\documentclass[twoside]{article}
\usepackage{fleqn,espcrc2}

\usepackage{graphicx}

\newcommand{\AmS}{{\protect\the\textfont2
  A\kern-.1667em\lower.5ex\hbox{M}\kern-.125emS}}

\title{Josephson Plasma Resonance in Solid and Glass Phases of
	Bi$_2$Sr$_2$CaCu$_2$O$_{8+\delta}$}

\author{
		Itsuhiro Kakeya
		\address{Institute of Materials Science, 
		University of Tsukuba, \\ 
        1-1-1 Ten-nodai, Tsukuba, Ibaraki 305-8573, Japan}$^\mathrm{b}$,
        Ryo Nakamura~$^\mathrm{a}$\thanks{Present address: Electrophotographic Laboratory, Canon Inc., 4202, Fukara, Susono, Shizuoka, 410-1196, Japan}, Tomoyuki Wada~$^\mathrm{a}$,
        and Kazuo Kadowaki~$^\mathrm{a}\!\!$
        \address{CREST, Japan Science and Technology Cooperation}}
       
\begin{document}

\begin{abstract}
Vortex matter phases and phase transitions are investigated by means of Josephson plasma resonance in under-doped Bi$_2$Sr$_2$CaCu$_2$O$_{8+\delta}$ single crystals in a microwave frequency range between 19 and 70 GHz. 
Accompanied by the vortex lattice melting transition, a jump of the interlayer phase coherence extracted from the field dependence of the plasma frequency was observed.  
In the solid phase, the interlayer coherence little depends on field at a temperature region well below $T_c$ while it gradually decreases as field increases toward the melting line up to just below $T_c$.  
As a result, 
the magnitude of the jump decreases with increasing temperature and 
is gradually lost in the vicinity of $T_c$.  
This indicates that the vortex lines formed in the vortex solid phase are thermally meandering and the phase transition becomes weak especially just below $T_c$. 
\vspace{1pc}
\end{abstract}

\maketitle
\begin{figure}[htb]
\begin{center}
\includegraphics[width=0.9\linewidth]{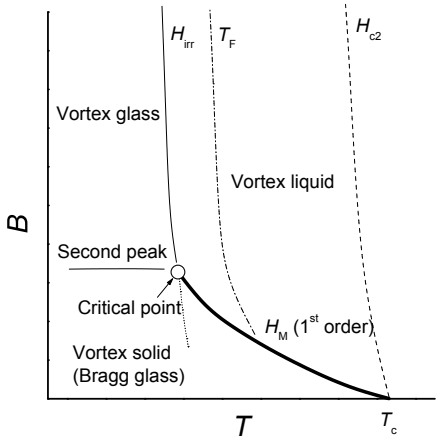}
\end{center}
\caption{
Schematic vortex phase diagram in high-$T_c$ superconductors 
with weak pinning in fields parallel to the $c$ axis.
}
\label{Phase}
\end{figure}

Studies of the mixed state in high-$T_c$ superconductors have brought 
about deeper understandings of the vortex matter in this decade.
Most significant notion here is in the existence of the first order 
vortex lattice melting phenomenon~\cite{Zel95} occurring at $H_M$
as shown in Fig.~\ref{Phase} schematically.
In high-$T_c$ superconductors, this lies well below the upper critical 
field $H_{c2}$ expected from the mean field approximation.
This drastic change of the phase diagram has led much interest especially in the vortex liquid state 
existing above the first order vortex lattice melting transition
(and the irreversibility line) below $H_{c2}$ line.
Since the macroscopic superconducting phase coherence is lost there, 
the vortex liquid state has been considered as a non-superconducting 
state having only the short range coherence.

In Bi$_2$Sr$_2$CaCu$_2$O$_{8+\delta}$ (BSCCO), 
the vortex lattice melting transition occurs 
in temperatures and magnetic fields
being easily accessible experimentally.
The first order character of the phase transition continues
up to the critical point, 
at which it changes the character to the second order phase transition
and continues to the irreversibility line and the horizontal line 
so called the second magnetization peak~\cite{Kha96}
as shown schematically in Fig.~\ref{Phase}.
So far, most of the Josephson plasma resonance (JPR) have been done 
in the vortex liquid phase except little study in the other vortex phases.
JPR can provide
quantitative information on the interlayer coherence between
adjacent superconducting CuO$_2$ layers in highly anisotropic 
superconductor such as BSCCO.
The Josephson plasma frequency 
in a finite field and a temperature is expressed as
\begin{equation}
\omega_p^2(H,T)= \omega_p^2(T) \langle \cos \varphi_{n,n+1} \rangle(H,T),
\label{eq:omega}
\end{equation}
where $\varphi_{n,n+1}$ is the gauge invariant phase difference 
between two adjacent $n$-th and $n+1$-th layers,
and $\langle \cdots \rangle$ means the thermal and the spatial average.
Here, $\omega_p(T) \equiv \sqrt{4 e d j_c(T)/\epsilon h}$ is the zero-field 
plasma frequency with $d$, $j_c(T)$, and $\epsilon$ being 
the distance between adjacent layers, 
the temperature dependent critical current along the $c$ axis,
and the high frequency dielectric constant, respectively~\cite{Bul95}.
Since the interlayer phase coherence $\langle \cos \varphi_{n,n+1} \rangle (H,T)$ directly indicates the $c$ axis correlations of pancakes, 
it is of considerable interest in studying the different vortex states
discriminated by the various phase transition 
lines by means of JPR~\cite{Mat95,Hana97,Kake00}.

In Ref. \cite{Shiba99,Gai00}, 
abrupt jump of the plasma frequency accompanied with
the melting transition is reported.
Their result indicates that 
the clear jump of the phase coherence continues up to $T_c$ 
and the jump suddenly disappears at $T_c$.
According to Blatter {\it et al.}, 
the first order melting transition is weaken near $T_c$~\cite{Bla96} 
and the symptoms have been found experimentally~\cite{Kad98}.
Quite recently, Koshelev and Bulaevskii also pointed out that
the interlayer coherence does not change considerably 
at the melting transition in low fields (in the vicinity of $T_c$)
because of strong meandering of the vortex lines~\cite{Kosh00}. 

In this paper, we discuss the melting transition quantitatively 
and the vortex state especially for the vortex solid phase below 
the melting transition in terms of the interlayer coherence 
extracted from JPR measurements as functions of microwave frequency, 
temperature, and magnetic field.
As consequences, the change of interlayer coherence at the transition 
is quite small especially in the vicinity of $T_c$, 
where the vortex lines are not straight but 
meandering strongly even in the vortex solid state.

JPR measurements were performed in a microwave frequency range between
19 and 70 GHz in an under-doped BSCCO single crystal with $T_c$ of 72.4 K.
Frequency-stabilized microwave was generated by a signal swept generator
or Gunn oscillators and the resonance was detected either by sweeping external magnetic field at a fixed temperature
or by sweeping temperature at a fixed field.
Details of the experimental setup are described in Ref.~\cite{Kake98}

\begin{figure}[htb]
\begin{center}
\includegraphics[width=0.9\linewidth]{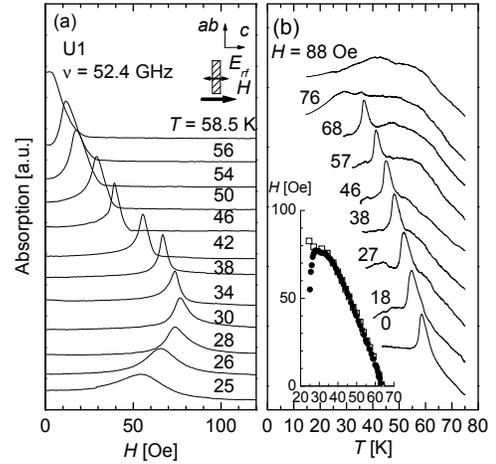}
\end{center}
\caption{
A set of JPR lines observed at 52.4 GHz at various temperatures 
obtained by sweeping field (a) 
and at various fields by sweeping temperature (b).
In the inset of (b), 
the resonance positions obtained by sweeping field and temperature 
are plotted as solid and open symbols, respectively.
These agree well except for temperatures below 35 K.
It denotes that these curves in (a) and (b) were obtained in increasing $H$ and $T$,
and no hysteresis due to sweeping $H$ and $T$ was found
except when field was sweptbelow 30 K.
}
\label{fig:line}
\end{figure}

Figure \ref{fig:line}(a) 
represents typical resonance curves obtained by sweeping magnetic field.
The resonance appears from 25 K in a finite field, 
the resonance field shows maximum around 30 K, 
and it disappears above 59 K,
where the zero field plasma frequency $\omega_p(T)$ 
coincides with the incidental microwave frequency of 52.4 GHz.
At all frequencies below 64 GHz, 
we observed similar feature of the resonance:
the resonance field once increase and decrease as temperature increases,
and the resonance disappears below $T_c$.
It is noted that the resonance becomes hysteretic and
quickly moves towards low field as temperature is decreased below 30 K,
and it is no longer observable below 20 K.
As clearly seen in Fig.~\ref{fig:line}(a), 
 It is also noticeable that
the resonance line width which becomes larger quickly 
below the temperature at which the resonance field is maximum.
By sweeping temperature on the contrary, 
a sharp resonance is observed 
below 68 Oe as shown in Fig.~\ref{fig:line}(b). 

\begin{figure}
\begin{center}
\includegraphics[width=0.9\linewidth]{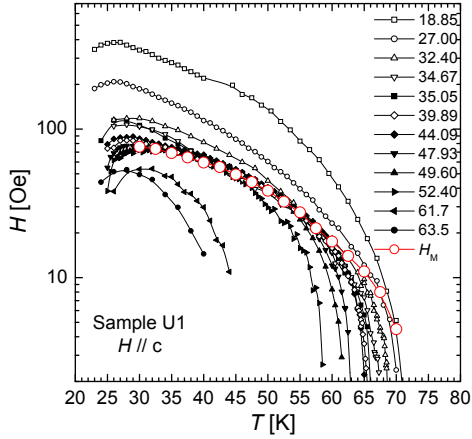}
\end{center}
\caption{
Temperature dependence of the resonance field $H_{res}$ for frequencies
between 18.85 and 63.5 GHz.
Small symbols indicate observed resonance fields and 
large open circle indicates the vortex lattice melting transition observed
by a SQUID magnetometer.
}
\label{plot:H-T}
\end{figure}

In Fig.~\ref{plot:H-T}, 
temperature dependences of the resonance field $H_{res}$ 
at 12 frequencies between 18 and 64 GHz together with
temperature dependence of the vortex lattice melting field ($H_M$)
obtained by the magnetization measurements by a SQUID magnetometer 
in the same crystal are plotted.
$H_{res}$ at a given temperature decreases with increasing frequency, 
and the resonance is observed not only in the vortex liquid state but also
in the vortex solid state below $H_M$.
The temperature dependence of $H_{res}$ below $H_M$ is similar to the
one in the liquid state: a broad maximum around 30 K 
and $H_{res}$ decrease toward $T_c$.
Below 30 K, $H_{res}$ decreases toward low temperature at all frequencies 
even below $H_M$.
As shown in the inset of Fig.~\ref{fig:line}(b),
the resonance positions obtained by sweeping both magnetic field
and temperature coincide above the maximum.
This means that the thermal equilibrium state of vortices is realized
even in the solid phase above the maximum by sweeping magnetic field,
whereas it is not below the maximum.
In previous studies, this inequilibrium state 
has been explained due to a pinning effect in the vortex glass state
because the maximum of $H_{res}$ almost coincides with $H_{irr}$
shown in Fig.~\ref{Phase}. 
Although this explanation may be valid for the resonance above $H_M$, 
the behaviors of the resonance in the solid phase implies that 
there is also a phase boundary 
similar to $H_{irr}$ line even below $H_M$ indicated as a dotted line
in Fig.~\ref{Phase}.

In our previous publication~\cite{Kad99}
the temperature dependence of JPR in zero magnetic field $\omega_p(T)$ was 
explained by the two fluid model very well as 
\begin{equation}
\omega_p^2(T)
	=\frac{\omega_p^2(0)}{2}
		\Bigl[ 
		1-\tilde{\tau}^{-2}
		+\sqrt{(1+\tilde{\tau}^{-2})^2-4\tilde{\tau}^{-2}(T/T_c)^4} 
		\Bigr],
\label{eq:f-T}
\end{equation}
where $\tilde{\tau}=\tau\omega_p(0)$ is the reduced scattering rate.
Using Eq. (\ref{eq:omega}) and Eq. (\ref{eq:f-T}), 
we can derive the interlayer coherence without ambiguity.
The normalized field dependence of the interlayer phase coherence 
$\omega_p^2(H,T)/\omega_p^2(T)$ at $H_M$ is plotted as a function of field
in Fig.~\ref{omega-H,T}. 
It is clearly observed that the interlayer phase coherence abruptly jumps 
in the vicinity of $H_M$ with decreasing magnetic field at low temperatres
below 55 K.
This jump can be attributed to the first order phase transition from liquid to
solid as reported previously~\cite{Shiba99}.
Although the magnitude of this jump is clearly noticeable,
it amounts only to 0.65 at most 
and becomes even smaller at higher temperatures.
Such a relatively small jump of the phase coherence at $H_M$ is partly caused
by the precursor reduction of the phase coherence already 
beginning well below $H_M$,
indicating that the strong thermal phase fluctuations
dominate in wide temperature region below $T_c$.
The interlayer coherence continues to survive even above $H_M$ 
and seems to follow the $1/H$ dependence very well in good accordance 
with the theoretical prediction~\cite{Kos96}.

The magnitude of the jump of the interlayer coherence 
decreases as temperature increases,
the sharp jump gradually disappears and turns to a broad increase above 65 K.
In terms of the interlayer coherence, 
this result is interpreted that the vortex lattice melting transition 
becomes weaker as temperature increases and finally looses 
one of the characteristics of the first order phase transition 
in temperatures considerably below $T_c$ (5--7 K).
This is in contrast to the results in Ref. \cite{Shiba99},
where the change of interlayer coherence accompanied by 
the first order transition seems not to depend on temperature.
Koshelev and Bulaevskii have derived a relation between 
plasma frequencies and the vortex wandering length $r_w$
defined as mean deviation of two pancake vortices of a vortex line
at neighboring layers, 
which does not depend on the wandering mechanism of vortex lines~\cite{Kosh00}.
They conclude that in the low field region of $B < 20$ G
the wandering length of the vortex lines is comparable with 
the Josephson length $\lambda_J$
and the interlayer coherence does not change considerably 
at the melting transition.
Here, $\lambda_J$ is given by the anisotropy parameter
$\gamma$ and the interlayer spacing $s$ as 
$\lambda_J \equiv \gamma s $.
This suggestion is consistent with our experimental results
in a sense that the abrupt jump at $T_M$ in the vicinity of $T_c$
is hardly visible due parhaps to the weakening of the melting transition 
caused by the strong vortex fluctuations even in the vortex solid state.

Finally, we mention about the line-shape of the resonance in the solid phase.
As shown in Fig.~\ref{fig:line}, 
the resonance line observed in the liquid state (between 35 and 46 K) is sharp 
and has a slight tail at higher field side, 
whereas the line in the glass and the solid phases is relatively broad 
and has a tail at lower field side.
The line-shape in the liquid state has been explained by the dispersion relation of 
the longitudinal Josephson plasma mode~\cite{Kad97}.
On the contrary to this, non-uniform field distribution---interlayer phase difference 
$\varphi_{n,n+1}$--- 
over a layer may be generated in the solid and the glass phases, 
since pancake vortices are localized because of formation of the vortex lines.
As a result, $\varphi_{n,n+1}$ has relatively large inhomogenity over a layer,
from which the low field tail and broadening of the line could be explained.

In summary, the vortex lattice melting transition is studied quantitatively 
in terms of interlayer coherence by the Josephson plasma resonance.
The transition becomes weaker at higher temperatures 
and the jump of the interlayer coherence remarkably diminishes near $T_c$.
This result suggest that the fluctuations of the vortices 
even in the vortex solid phase 
is considerably strong and 
the melting transition may turn from the first to the second order.

This work is partly supported by the REIMEI Research Resources of 
Japan Atomic Energy Research Institute.

\begin{figure}
\begin{center}
\includegraphics[width=0.9\linewidth]{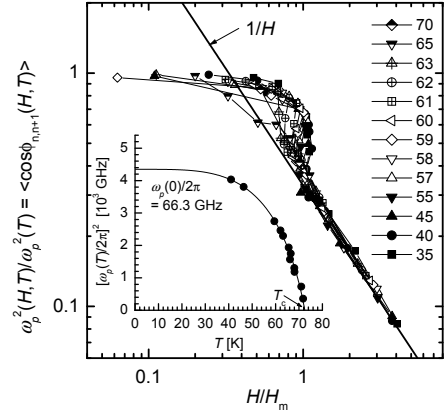}
\end{center}
\caption{
The interlayer coherence $\langle \cos \varphi_{n,n+1}\rangle (H,T)$ 
plotted as a function of magnetic field normalized by $H_M$.
Magnitude of the jump of $\langle \cos \varphi_{n,n+1}\rangle(H,T) $ at the vortex lattice melting transition is larger at lower temperatures.
Thick line follows $\langle \cos \varphi_{n,n+1}\rangle (H,T) \propto H^{-1}$.
The inset shows the temperature dependence of the zero field plasma frequency.
Solid symbols indicate experimental results and
solid lines were obtained by fitting Eq. (\ref{eq:f-T})
to experimental results.
The agreement between experiment and the model is very good.
}
\label{omega-H,T}
\end{figure}

\end{document}